\documentclass[aps,pra%
	,amsmath%
	,reprint%
	]{revtex4-2}

\usepackage{graphicx}
\newcommand{\bvec}[1]{\boldsymbol{#1}} 
\newcommand{\set}[1]{\left\{ #1 \right\}}

\newcommand{\rotmat}{R}
\newcommand{\id}{I_3}
\newcommand{\R}[2]{R\left(#1, #2\right)}
\newcommand{\Rmat}[2]{R_{#1}\left(#2\right)}

\newcommand{\Bs}{\bvec{B}_{\rm s}}
\newcommand{\Bp}{\bvec{B}_{\rm p}}
\newcommand{\bs}{\bvec{b}_{\rm s}}
\newcommand{\bp}{\bvec{b}_{\rm p}}

\newcommand{\Floquet}{F}
\newcommand{\Fs}{\Floquet_3^{({\rm s})}}

\begin{document}

\title{Which Adiabatic Cycles Alter an Action Variable?}

\author{Atushi Tanaka}
\homepage[]{https://orcid.org/0000-0002-1070-6336}
\affiliation{Department of Physics, Tokyo Metropolitan University,
  Hachioji, Tokyo 192-0397, Japan}
\author{Taksu Cheon}
\affiliation{Laboratory of Physics, Kochi University of Technology,
  Tosa Yamada, Kochi 782-8502, Japan}

\date{2026-10-19}

\begin{abstract}
  Action variables are adiabatic invariants: their values are preserved during adiabatic time evolutions in completely integrable classical Hamiltonian systems. Nevertheless, an adiabatic cycle can alter
  how an action variable depends on
  phase-space variables that are independent of adiabatic parameters.
  Such a change, which we call action variable holonomy,
  induces adiabatic torus pumping in phase space.
  For a family of completely integrable kicked-top systems, a homotopy classification of adiabatic cycles identifies which adiabatic cycles induce action variable holonomy and adiabatic torus pumping. A physical example is also provided.
\end{abstract}


\maketitle

\section{Introduction}
\label{sec:introduction}

Action variables are fundamental for classical integrable systems.
Their first role is
to label invariant tori: each set of their values specifies
a torus that is invariant under time evolution.
Their second role is to provide a reference for describing time evolution. In particular, under an adiabatic perturbation of a completely integrable system, action variables are adiabatic invariants, unlike other conserved quantities.

There are two aspects to the behavior of an action variable under an adiabatic change in external parameters.
First, its value remains unchanged by virtue of adiabatic invariance.
Second, because the action variable is also a function of canonical variables independent of the parameters, its functional form changes under parameter deformation.
The latter represents the continuation of coordinates adapted to each instantaneous integrable system.
The cumulative effect of this continuation can be nontrivial, as suggested by the notion of analytic continuation.

In fact,
for a harmonic chain with clamps, it has been reported that an action variable,
viewed as a function, changes after the parameters are varied once along a closed
path~\cite{Tanaka-PLA-460-128621}.
By analogy with geometric phases~\cite{Berry-PRSLA-392-45,Bohm-GPQS-2003,Chruscinski-GPCQM-2004}, we call this change action variable holonomy.

One notable consequence of action variable holonomy is found in the motion of a torus during an adiabatic cycle. While the value of the action variable is preserved throughout the cycle,
a torus initially specified by this value is carried, at the end of the cycle, to another torus specified by the same value but a different functional form of the action variable.
We call this phenomenon adiabatic torus pumping in phase space.

We also note that the absence of nontrivial holonomy of the action variables is required for the Hannay angle~\cite{Hannay-JPA-18-221} to be well defined, and the above example shows that this requirement is not automatically satisfied.
A related issue concerning the definition of classical adiabatic angles is discussed in Ref.~\cite{Montgomery-CMP-120-269}.

The existence of action variable holonomy raise the question of which adiabatic cycles induce action variable holonomy. Intuitively, small cycles are not expected to change action variables~\cite{Montgomery-CMP-120-269}.
In other words,
action variable holonomy should require a ``large'' cycle.

In this paper, we consider a Floquet system as an example exhibiting action variable holonomy.
Its adiabatic parameter space is simple yet admits diverse paths.
This allows us to determine which cycles induce action variable holonomy.
We further recast this answer in terms of a homotopy classification of adiabatic cycles.

The paper is organized as follows.
Section~\ref{sec:model} introduces
classical kicked tops, a family of completely integrable classical Floquet systems.
For fixed system parameters, this system is equivalent to a classical spin in a static magnetic field~\cite{Berry-NATOASI-144-267,Chruscinski-GPCQM-2004}. We use this equivalence to obtain the action variable.
Section~\ref{sec:adevol} discusses the adiabatic time evolution of this system. We introduce a parametrization for describing the topological properties of adiabatic cycles and use it to identify the cycles that induce action variable holonomy.
We also describe the resulting adiabatic torus pumping in phase space.
Section~\ref{sec:homotopy} expresses the condition for action variable holonomy obtained in the preceding section in a parametrization-independent form through a homotopy classification of adiabatic cycles.
Section~\ref{sec:implementation} presents a physical realization of action variable holonomy using a classical spin subject to a static magnetic field and a time-periodic pulsed magnetic field.
Sections~\ref{sec:discussion} and \ref{sec:summary} present the discussion and summary, respectively.
The appendices provide results needed in Sec.~\ref{sec:implementation}.

\section{A family of completely integrable classical Floquet systems}
\label{sec:model}

We consider a completely integrable classical kicked top.
As explained in Sec.~\ref{sec:implementation}, this system can be implemented as a Floquet system~\cite{Haake-ZPB-65-381}.
It is a discrete-time dynamical system described by a linear map acting on the classical angular momentum $\bvec{S}$:
\begin{equation}
  \label{eq:map}
  \bvec{S}''
  =
  \rotmat
  \bvec{S}'
  ,
\end{equation}
where $\rotmat$ is a rotation matrix.
That is, $\rotmat$ is an element of $\textrm{SO}(3)$, the set of $3\times3$ orthogonal matrices with determinant $1$.
Since $S^2=\bvec{S}\cdot\bvec{S}$ is conserved, we fix its value throughout.
Under this condition, the system can be regarded as a symplectic map in two variables~\cite{Scharf-JPA-21-2007}.

In this section, we outline the properties of the map~\eqref{eq:map} for a fixed $\rotmat$.
This prepares the discussion of adiabatic time evolution in the next section.

The rotation matrix $\rotmat$ can be specified by the rotation axis $\bvec{n}$, a unit vector, and the rotation angle $\theta$:
\begin{align}
  \label{eq:R_by_ntheta}
    \R{\bvec{n}}{\theta}
  &
    =
    \begin{bmatrix}
      \bvec{n}_1&\bvec{n}_2&\bvec{n}
    \end{bmatrix}
    \begin{bmatrix}
      \cos\theta& -\sin\theta&0\\
      \sin\theta& \cos\theta&0\\
      0& 0& 1\\
    \end{bmatrix}%
    \begin{bmatrix}
      {}^{\rm t}\bvec{n}_1\\{}{}^{\rm t}\bvec{n}_2\\{}{}^{\rm t}\bvec{n}
    \end{bmatrix}%
\end{align}
where $\bvec{n}_1$ is a unit vector orthogonal to $\bvec{n}$, and $\bvec{n}_2=\bvec{n}\times\bvec{n}_1$.
Eq.~\eqref{eq:R_by_ntheta} is uniquely determined once $\bvec{n}$ is fixed and is therefore independent of the choice of $\bvec{n}_1$.

The map~\eqref{eq:map} induced by the rotation matrix $\R{\bvec{n}}{\theta}$ specified by $\bvec{n}$ and $\theta$ can also be regarded as the Poincar\'e map for a spin in a static magnetic field along $\bvec{n}$.
Using this correspondence, we can determine the conserved quantity and the action variable of the map
(cf. Refs.~\cite{Berry-NATOASI-144-267,Chruscinski-GPCQM-2004}).

Indeed, for the map~\eqref{eq:map},
the spin component along the rotation axis $\bvec{n}$,
\begin{align}
  \label{eq:I_def}
  I(\bvec{n})=\bvec{n}\cdot\bvec{S}
\end{align}
is conserved.
This quantity is independent of the conserved quantity $S^2$.
Therefore, the system is completely integrable~\cite{Maeda-PJAA-63-198,Bruschi-PhysicaD-49-273}.

Moreover, $I(\bvec{n})$ is also an action variable~\cite{Berry-NATOASI-144-267,Chruscinski-GPCQM-2004}.
It is obtained by the standard procedure for constructing an action variable from a conserved quantity~\cite{Arnold-MethodOfClassical}.
The value of $I(\bvec{n})$ specifies a one-dimensional torus (circle) in phase space.
Thus, $I(\bvec{n})$ serves as a coordinate specifying the torus.

We recall how the adiabatic time evolution of the map~\eqref{eq:map} changes the position of tori in phase space. Consider a torus initially specified by $I(\bvec{n})=I_0$. As $\rotmat$ changes adiabatically, the rotation axis $\bvec{n}$ changes, and hence the functional form of $I(\bvec{n})$ changes. On the other hand, its value remains invariant. Therefore, the torus is continuously specified by $I(\bvec{n})=I_0$ throughout the adiabatic evolution, while its position changes with $\bvec{n}$.

Note that, when $\rotmat$ is the identity matrix $\id$, $I(\bvec{n})$ cannot be defined because the rotation axis $\bvec{n}$ has no meaning.
Equivalently, the map~\eqref{eq:map} conserves every physical quantity.

\section{Action variable holonomy}
\label{sec:adevol}

In the previous section, we obtained the action variable $I(\bvec{n})$ (Eq.~\eqref{eq:I_def}), which is an adiabatic invariant of the linear map~\eqref{eq:map}. Here we examine how the functional form of $I(\bvec{n})$ as a function of the phase-space variable $\bvec{S}$ changes with $\rotmat$ in Eq.~\eqref{eq:map}.

We mainly consider the case in which $\rotmat$
slowly
traces a closed path, namely, an adiabatic cycle.
Open paths are discussed at the end of this section.
In discussing the adiabatic time evolution, however, we restrict ourselves to paths of $\rotmat$ that do not pass through the identity matrix $\id$.
This is because the action variable $I(\bvec{n})$ cannot be continued continuously along a path passing through $\id$.

When discussing adiabatic time evolution, it is inconvenient to parametrize $\rotmat$ by $(\bvec{n},\theta)$.
For a rotation matrix $\rotmat$, $(\bvec{n}, \theta)$ and $(-\bvec{n}, -\theta)$ represent the same rotation matrix.
Moreover, $\theta$ is periodic.
It is cumbersome to discuss the geometry of a path of $\rotmat$ while keeping track of these redundancies.

We therefore use the angle vector $\bvec{\theta}$.
Its direction and magnitude specify the rotation axis and rotation angle, respectively~\cite{thetavector,Johansson-PRL-92-060406,Khatua-PRB-105-184401}.
For a given $\bvec{\theta}$, there are two choices of the rotation axis:
\begin{align}
  \label{eq:n_pm_by_thetavec}
  \bvec{n}_{\pm}(\bvec{\theta})=\pm\frac{\bvec{\theta}}{|\bvec{\theta}|}
  .
\end{align}
The rotation angles corresponding to these axes are $\pm|\bvec{\theta}|$ modulo $2\pi$.
We define the rotation matrix corresponding to $\bvec{\theta}=\bvec{0}$ to be the identity matrix.

Conversely, we describe how to determine the angle vector $\bvec{\theta}$ from a given rotation axis $\bvec{n}$ and rotation angle $\theta$.
First, let $\theta'=\theta - 2\pi\lfloor \theta/(2\pi)\rfloor$.
For $\theta'\ne\pi$,
\begin{align}
  \label{eq:thetavec_by_theta_n}
  \bvec{\theta}
  =
  \begin{cases}
    \theta'\bvec{n}&\text{for $0\le \theta' < \pi$}\\
    -(2\pi-\theta')\bvec{n}&\text{for $\pi < \theta' < 2\pi$}
  \end{cases}
  .
\end{align}
This definition gives $|\bvec{\theta}|<\pi$.
For $\theta'=\pi$, on the other hand, $\bvec{\theta}=\pm\theta'\bvec{n}$ specifies the same rotation matrix.
This corresponds to the fact that rotations by angle $\pi$ about the axes $\pm\bvec{n}$ are identical.

Thus, the range $B$ of the angle vector $\bvec{\theta}$ may be restricted to the three-dimensional ball of radius $\pi$.
Except on the surface of $B$, $\bvec{\theta}$ and $\rotmat$ (an element of ${\rm SO}(3)$) are in one-to-one correspondence (see, e.g., \S~1.3.4 of Ref.~\cite{Hall-LieGroupsLieAlgebrasAndRepresentations2nd-2015}).
On the surface of $B$, where $|\bvec{\theta}|=\pi$, an antipodal pair $\pm\bvec{\theta}$ specifies the same rotation matrix.
Therefore, when a continuous path crosses the surface of $B$ at $\bvec{\theta}_0$, its representative in the ball moves immediately to the antipodal point $-\bvec{\theta}_0$.

Furthermore, we exclude $\bvec{0}$ from $B$.
As noted above, this is necessary because the point corresponding to $\rotmat=\id$ must be avoided when discussing adiabatic time evolution.

To clarify the parametric evolution of the action variable $I(\bvec{n})$ (Eq.~\eqref{eq:I_def}) as a function on phase space along a closed path $C$ traced by $\rotmat$, we first examine the parametric evolution of $\bvec{n}$.
For this purpose, we examine its relation to the two functions $\bvec{n}_{\pm}(\bvec{\theta})$ (Eq.~\eqref{eq:n_pm_by_thetavec}).
Let $\bvec{n}_0$ be the value of $\bvec{n}$ at a point $\bvec{\theta}_0$ on $C$, and let $\bvec{n}_1$ be its value when the same point $\bvec{\theta}_0$ is reached after one traversal of $C$.
In the following, we compare $\bvec{n}_0$ and $\bvec{n}_1$.

Suppose that $C$ lies entirely inside $B$, i.e., $0<|\bvec{\theta}|<\pi$ along $C$.
Then the parametric evolution of $\bvec{n}$ follows one of the two single-valued functions $\bvec{n}_{\pm}(\bvec{\theta})$ throughout the cycle, and hence $\bvec{n}_1=\bvec{n}_0$.
The branch is determined by whether $\bvec{n}_0=\bvec{n}_+(\bvec{\theta}_0)$ or $\bvec{n}_-(\bvec{\theta}_0)$.

\begin{figure}[tb]
  \centering
  \includegraphics[%
	width=0.3\textwidth%
        ]{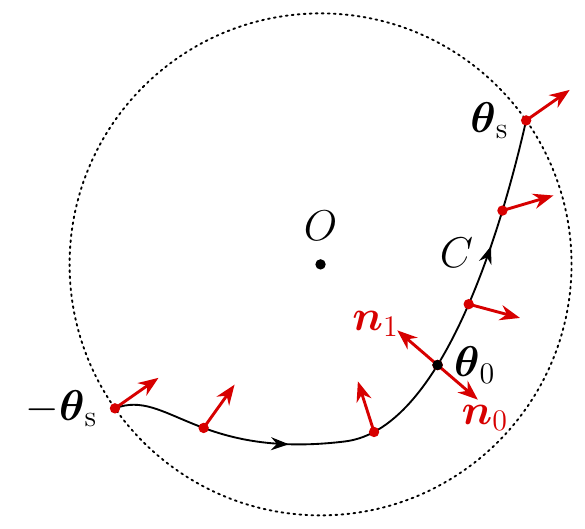} 
  \caption{%
    A closed path $C$ (solid line) that crosses the surface of the ball $B$ once, and the parametric evolution of $\bvec{n}$ (arrows) along $C$.
    The intersection of the surface of $B$ with the $(\theta_x,\theta_y)$ plane is shown by the dotted line.
    Starting from $\bvec{\theta}_0$, $C$ reaches the point $\bvec{\theta}_{\textrm{s}}$ on the surface of $B$. The points $\bvec{\theta}_{\textrm{s}}$ and $-\bvec{\theta}_{\textrm{s}}$ are identified because they correspond to the same rotation matrix.
    It then returns to $\bvec{\theta}_0$, forming a closed path with $\nu(C)=1$.
    Let $\bvec{n}_0=\bvec{n}_+(\bvec{\theta}_0)$ at the starting point.
    Continuity of $\bvec{n}$ along $C$ requires a switch from $\bvec{n}_+$ to $\bvec{n}_-$ at the surface crossing, giving $\bvec{n}_1=\bvec{n}_-(\bvec{\theta}_0)=-\bvec{n}_0$ at the endpoint.
    Thus, the parametric evolution along $C$ reverses $\bvec{n}$.
  }
  \label{fig:n_evol_gamma}
\end{figure}

When $C$ crosses the surface of $B$, however, $\bvec{n}$ need not return to $\bvec{n}_0$ (Fig.~\ref{fig:n_evol_gamma}). At a surface crossing, antipodal points are identified in the ball representation, so that the functions $\bvec{n}_{\pm}(\bvec{\theta})$ change discontinuously. Since the functional form of the action variable $I(\bvec{n})$~(Eq.~\eqref{eq:I_def}) is required to vary continuously along $C$, the parametric evolution of $\bvec{n}$ must remain continuous.
Therefore, the branch followed by the parametric evolution switches between $\bvec{n}_{+}$ and $\bvec{n}_{-}$ at each surface crossing.

In particular, a single crossing gives $\bvec{n}_1=-\bvec{n}_0$ (see Fig.~\ref{fig:n_evol_gamma}), whereas merely touching the surface does not switch the branch and hence does not reverse $\bvec{n}$. If $\nu(C)$ denotes the number of surface crossings, the general result is
\begin{equation}
  \bvec{n}_1=(-1)^{\nu(C)}\bvec{n}_0.
\end{equation}

This result is directly reflected in the parametric evolution of the action variable $I(\bvec{n})$.
Using the result above, we obtain
\begin{equation}
  \label{eq:I1_by_I0}
  I(\bvec{n}_1)=(-1)^{\nu(C)}I(\bvec{n}_0)
\end{equation}
at the endpoint of the cycle. Thus, odd $\nu(C)$ implies holonomy of the action variable.

\begin{figure}[htb]
  \centering
  \includegraphics[%
	width=0.45\textwidth%
        ]{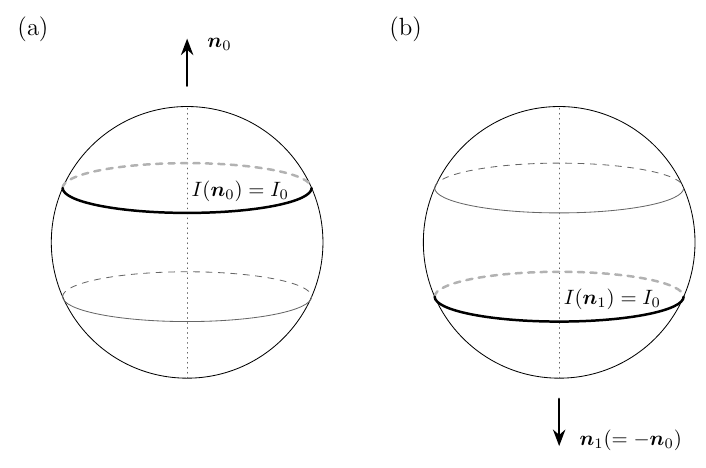}  
  \caption{%
    Adiabatic torus pumping in phase space.
    (a) At the starting point of the adiabatic cycle, let $\bvec{n}=\bvec{n}_0$.
    We focus on the one-dimensional torus (thick line) specified by $I(\bvec{n}_0)=I_0$.
    The torus corresponding to $I(\bvec{n}_0)=-I_0$ (thin line) is also shown;
    (b) At the endpoint of the adiabatic cycle, let $\bvec{n}=\bvec{n}_1$.
    By the adiabatic theorem, the torus $I(\bvec{n}_0)=I_0$ considered at the starting point is transferred to the torus corresponding to $I(\bvec{n}_1)=I_0$.
    When $(-1)^{\nu(C)}=-1$, we have $\bvec{n}_1=-\bvec{n}_0$, so the final torus does not return to the position of the initial torus (thin line), but instead becomes the torus on the ``opposite side'' (thick line).
  }
  \label{fig:torus_pump}
\end{figure}

We now show that, as a consequence of action variable holonomy, an adiabatic cycle $C$ induces adiabatic torus pumping in phase space.
Let us examine a torus specified by the value $I_0$ of the initial action variable $I(\bvec{n}_0)$.
The torus $I(\bvec{n}_0)= I_0$ is a circle on the sphere $\bvec{S}^2=S^2$ for which the component of $\bvec{S}$ along the
$\bvec{n}_0$
axis is $I_0$, with $-S \le I_0 \le S$ (Fig.~\ref{fig:torus_pump}~(a)).
Since $I(\bvec{n})= I_0$ holds throughout the adiabatic evolution, we obtain $I(\bvec{n}_1)=I_0$ at the end of $C$.
Because of the action variable holonomy~\eqref{eq:I1_by_I0}, we obtain $(-1)^{\nu(C)}I(\bvec{n}_0)=I_0$,
which tells us where the torus arrives at the end of $C$.
If $\nu(C)$ is even, the torus returns to its original position.
If $\nu(C)$ is odd, the torus is transferred to another torus specified by $I(\bvec{n}_0)= -I_0$ (Fig.~\ref{fig:torus_pump}~(b)).
For $I_0=0$, the torus returns to its original position regardless of the parity of $\nu(C)$.

Finally, we extend the discussion from closed paths to open paths.
Consider two open paths $C_0$ and $C_1$ with the same starting point and endpoint, and compare the two adiabatic time evolutions along them.
Specifically, for the same initial action variable, we examine whether the functional forms of the action variable at the endpoint are identical after the adiabatic time evolutions along $C_0$ and $C_1$.
This comparison is determined by the cycle $C=C_0\cdot C_1^{-1}$ obtained by following $C_0$ and then the reverse of $C_1$.
If the number $\nu(C)$ of crossings of the surface of $B$ is even, the action variable does not change sign. 
Thus, the two adiabatic time evolutions along $C_0$ and $C_1$ give the same result.
If $\nu(C)$ is odd, the action variables resulting from $C_0$ and $C_1$ have opposite signs. 
Similarly, the final tori reached along the adiabatic paths $C_0$ and $C_1$ are the same when $\nu(C)$ is even and lie on opposite sides when $\nu(C)$ is odd.
For $I_0=0$, however, the final torus is the same regardless of the parity of $\nu(C)$.

\section{Which Adiabatic Cycles Alter an Action Variable?}
\label{sec:homotopy}

So far, we have seen that an adiabatic cycle $C$ induces action variable holonomy when $\nu(C)$, the number of crossings of the surface of $B$, is odd.
However, this criterion is formulated in terms of the surface of $B$ and is therefore not manifestly intrinsic.
In this section, we show that it can be recast in terms of the homotopy classification~\cite{Mermin-RMP-51-591} of adiabatic cycles.

We identify closed paths that can be continuously deformed into one another. The set of equivalence classes $[C]$ under this relation is called the fundamental group $\pi_1(B)$ of the space $B$~\cite{Mermin-RMP-51-591}.

We now show that $\pi_1(B)=\set{[e],[\gamma]}$. Here, $e$ is a closed path contractible to a point, and $\gamma$ is a closed path connecting a point on the surface of $B$ to its antipodal point through the interior of $B$.
Note that antipodal points on the surface of $B$ are identified.

To this end, we introduce the space $B'$ obtained by adding $\bvec{0}$ to $B$ and determine its fundamental group.
From the correspondence between $\rotmat$ and $\bvec{\theta}$, ${\rm SO}(3)$ can be identified with $B'$. Since $\pi_1({\rm SO}(3))=\set{[e],[\gamma]}$, we have $\pi_1(B')=\set{[e],[\gamma]}$.

The class $[e]$ in $B$ is obtained from $[e]$ in $B'$ by removing the closed paths that pass through $\bvec{0}$.
Note that closed paths belonging to $[e]$ in $B$ can be continuously deformed into one another while avoiding the origin. The same holds for $[\gamma]$. Therefore, $\pi_1(B)=\set{[e],[\gamma]}$.

Next, we show that a path $C$ with even $\nu(C)$ is homotopic to $e$, whereas a path $C$ with odd $\nu(C)$ is homotopic to $\gamma$.
First, when a closed path $C$ satisfies $\nu(C)=0$, it lies entirely inside $B$ and is therefore homotopic to $e$. When $\nu(C)=1$, the points where $C$ passes through the ``surface'' of $B$ are $P_1$ and its antipodal point $\bar{P}_1$, so $C$ is homotopic to $\gamma$.

For $\nu(C)\ge2$, surface crossings can be removed in pairs by a continuous deformation.
For example, when
$\nu(C)=2$, $C$ passes through the surface of $B$ in the order $P_1$, $\bar{P}_1$, $P_2$, $\bar{P}_2$. By continuously deforming $C$ so that $P_1$ and $\bar{P}_2$ coincide, $C$ becomes tangent to the surface of $B$ at $P_1$ and ${P}_2$. Since $P_1$ and ${P}_2$ can then be continuously moved into the interior of $B$, $C$ is homotopic to $e$.

Therefore, a necessary and sufficient condition for holonomy of the action variable to occur is that a closed path $C$ be homotopic to $\gamma$.

\section{A physical example}
\label{sec:implementation}

We introduce a physical system that realizes representative adiabatic cycles of the map~(\ref{eq:map}), in particular a contractible cycle $e$ and a noncontractible cycle $\gamma$.

Here we consider a classical angular momentum $\bvec{S}$ with magnetic moment $\gamma_{S}\bvec{S}$, where $\gamma_{S}$ is the gyromagnetic ratio.
We apply a time-dependent magnetic field $\bvec{B}(t)$ consisting of a static field $\Bs$ and periodic pulses:
\begin{align}
 \label{eq:B_t}
  \bvec{B}(t)
  &
    = \Bs + \Bp\sum_{n=-\infty}^{\infty}\delta\left(\frac{t}{T}-n\right)
    .
\end{align}
Here, $\Bp$ specifies the direction and strength of the pulsed magnetic field, and $T$ is the pulse period.
This system is described by the nonautonomous Hamiltonian
\begin{align}
  \label{eq:Hkickedtop}
  H(t) = -\gamma_{S}\bvec{B}(t)\cdot\bvec{S}
\end{align}
with the corresponding equation of motion
$d\bvec{S}/dt = \gamma_{S}\bvec{S}\times\bvec{B}(t)$.
We note that a quantum counterpart of the kicked top described by Eq.~\eqref{eq:Hkickedtop}
has been examined in Refs~\cite{Cheon-EPL-85-20001,TANAKA-AP-85-1340}.
Its relation to the present work will be discussed Sec.~\ref{sec:discussion}.

We introduce the rotation angles induced by the static and pulsed magnetic fields over one period $T$ as
\begin{gather}
  \label{eq:mu_lambda_def}
  \mu\equiv-\gamma_{S}\left|\Bs\right| T
  ,\quad
  \lambda\equiv-\gamma_{S}\left|\Bp\right|T
  .
\end{gather}%

Since this is a periodically kicked system, we can explicitly obtain the Floquet matrix:
\begin{align}
  \label{eq:Floquet_def}
  \Floquet
  &
    =
    \R{\bs}{\mu}
    \R{\bp}{\lambda}
    ,
\end{align}
where $\bs$ and $\bp$ are unit vectors along $\Bs$ and $\Bp$, respectively.
The Poincar\'e map $\Floquet$,
defined from immediately before one magnetic pulse to immediately before the next,
gives the rotation matrix $\rotmat$ in the map~\eqref{eq:map}.
To realize an adiabatic cycle, the parameters of $\Floquet$ must be varied on a time scale much longer than the period $T$ of the time-dependent magnetic field~\eqref{eq:B_t}.

In the following, we keep the static magnetic field fixed and vary only the pulsed magnetic field slowly.
We therefore choose the direction of $\Bs$ as the $z$ axis and set $\bs=\bvec{e}_3$.
The direction of the pulsed magnetic field is specified by the polar coordinates $\varphi$ and $\beta$:
\begin{gather}
  \label{eq:bp_parametrized}
  \bp=\sin\beta(\cos\varphi\bvec{e}_1+\sin\varphi\bvec{e}_2)+\cos\beta\bvec{e}_3
  .
\end{gather}

To make the parameter dependence of $F$ explicit, we write $F$ as follows (see Appendix~\ref{sec:eq:F_into_F3}):
\begin{align}
  \label{eq:F_into_F3}
  F=R_3\left(\varphi+\frac{\pi}{2}\right)
  \Floquet_3(\mu,\beta,\lambda)
  \; {}^{\rm t}R_3\left(\varphi+\frac{\pi}{2}\right)
\end{align}
where $R_i(x)=\R{\bvec{e}_i}{x}$ and
\begin{align}
  \label{eq:F3_def}
  \Floquet_3(\mu,\beta,\lambda)
  = R_3(\mu)R_1(\beta)R_3(\lambda) \;{}^{\rm t} R_1(\beta)
  .
\end{align}
The information on the static and pulsed magnetic fields is encoded in $F$ through the parameters $\mu$, $\lambda$, $\beta$, and $\varphi$.
To avoid unnecessary case distinctions involving the trivial case $F=I_3$, we impose $0 < \beta < \pi$ and $\mu\ne 0 \mod 2\pi$.

The product $F$ of rotation matrices is itself a rotation matrix.
We denote its rotation axis by $\bvec{n}$ and its rotation angle by $\theta$, where $\theta$ satisfies (see Appendix~\ref{sec:Floquet_rotation_axis_and_angle})
\begin{align}
  \label{eq:theta_eq}
  \cos\frac{\theta}{2}
  =\cos\frac{\lambda}{2}\cos\frac{\mu}{2}
  -\cos\beta\sin\frac{\lambda}{2}\sin\frac{\mu}{2}
  .
\end{align}
The conditions imposed on $\beta$ and $\mu$ ensure that $\theta\ne0\mod 2\pi$.
Hence, $F$ is never the identity matrix, and its rotation axis is always well defined (see Appendix~\ref{sec:Floquet_rotation_axis_and_angle}):
\begin{align}
  \label{eq:n_def}
  \bvec{n}
  =
  R_3\left(\varphi+\frac{\mu+\pi}{2}\right)
  \begin{bmatrix}
    0\\ -\sin\alpha\\{}\cos\alpha
  \end{bmatrix}
  ,
\end{align}
where $\alpha$ satisfies
\begin{align}
  \label{eq:alpha_def}
  \cos\alpha
  &= \frac{1}{\sin({\theta}/{2})}%
  \left(\cos\frac{\lambda}{2}\sin\frac{\mu}{2}
  +\cos\beta\sin\frac{\lambda}{2}\cos\frac{\mu}{2}\right)
  \nonumber\\
  \sin\alpha
  &= \frac{1}{\sin({\theta}/{2})}
  \sin\beta\sin\frac{\lambda}{2}
  .
\end{align}

\subsection{Cycle $C_1$ with a rotating pulsed magnetic field}

In the adiabatic cycle $C_1$ introduced here, we keep the static magnetic field $\Bs$ fixed and rotate the direction of the pulsed magnetic field $\Bp$ once around $\Bs$.
That is, we keep $\mu$, $\beta$, and $\lambda$ fixed and vary $\varphi$ by $2\pi$ (see Eqs.~\eqref{eq:mu_lambda_def} and~\eqref{eq:bp_parametrized}).
To exclude trivial cases, we impose $\lambda\ne 0\mod 2\pi$, where $\lambda$ corresponds to the strength of the pulsed magnetic field.
As stated above, we also assume $0<\beta<\pi$ and $\mu\ne 0\mod2\pi$.

Along this cycle, the factor $\Floquet_3(\mu,\beta,\lambda)$ of $F$ (Eq.~\eqref{eq:F_into_F3}) remains unchanged.
Hence, $\theta$ and $\alpha$ are independent of $\varphi$ and remain constant.
In particular, the conditions on $\beta$ and $\mu$ imply $\theta\ne 0\mod2\pi$ (see Eq.~\eqref{eq:theta_eq}).
The condition on $\lambda$ also implies $\alpha\ne 0\mod\pi$ (see Eq.~\eqref{eq:alpha_def}).

On the other hand, the rotation axis $\bvec{n}$ makes one revolution around the $z$ axis along the adiabatic cycle $C_1$ (see Eq.~\eqref{eq:n_def}).
Therefore, the action variable also returns to its original functional form at the end of $C_1$.

This result is consistent with the fact that the trajectory of the Floquet matrix along $C_1$ is homotopic to a point.
To verify this, we examine the trajectory of the angle vector $\bvec{\theta}$.
Defining $\bvec{\theta}$ according to Eq.~\eqref{eq:thetavec_by_theta_n}, we can show that $\bvec{\theta}$ traces a closed path along $C_1$ and that this path is contractible to a point.
First, $|\bvec{\theta}|$ remains constant along $C_1$.
For $|\bvec{\theta}| < \pi$, the direction of $\bvec{\theta}$ makes one revolution, so the trajectory of $\bvec{\theta}$ is closed.
This closed path does not cross the surface of $B$.
Thus, $\nu(C_1)=0$, and the path is homotopic to a point.
For $|\bvec{\theta}|=\pi$, on the other hand, the one-to-one correspondence between the rotation matrix and $\bvec{\theta}$ breaks down.
We therefore consider directly the trajectory of the rotation matrix.
This trajectory is homotopic to $\gamma^2$ and is thus contractible to a point.

\subsection{Cycle $C_2$ with monotonically varying pulsed-field strength}
We introduce a cycle $C_2$ in which the strength of the pulsed magnetic field $\Bp$ varies monotonically.
Along $C_2$, the direction $\bp$ of the pulsed magnetic field and the static magnetic field $\Bs$ are kept fixed.
Thus, $\mu$ and $\beta$ also remain constant.

The strength $|\Bp|$ of the pulsed magnetic field is proportional to the rotation angle $\lambda$ induced by a single pulse (Eq.~\eqref{eq:mu_lambda_def}).
Since the Floquet matrix $\Floquet$ is periodic in $\lambda$, the cycle $C_2$ is obtained by monotonically increasing or decreasing $|\Bp|$ until $\lambda$ changes by $2\pi$.
We note that, along $C_2$, the pulsed magnetic field $\Bp$ traces an open path in the external-parameter space, whereas the corresponding Floquet matrix $\Floquet$, or equivalently $\rotmat$ in Eq.~\eqref{eq:map}, traces a closed path.

The cycle $C_2$ is an example of a path homotopic to $\gamma$.
To show this, we examine the trajectories of the rotation angle $\theta$ and the rotation axis $\bvec{n}$ along $C_2$.

First, as in the case of $C_1$, the rotation angle $\theta$ satisfies Eq.~\eqref{eq:theta_eq}.
From this equation, we obtain
\begin{align}
  \label{eq:theta_holonomy}
  \theta(\mu,\beta,\lambda\pm2\pi)
  = 2\pi - \theta(\mu,\beta,\lambda)
\end{align}
(see Appendix~\ref{sec:theta_alpha_vs_lambda}).
On the other hand, Eq.~\eqref{eq:n_def} gives the rotation axis $\bvec{n}$ as
\begin{align}
  \bvec{n}
  =
  \begin{bmatrix}
    \cos\left(\varphi+\frac{\mu}{2}\right)\sin\alpha
    \\
    \sin\left(\varphi+\frac{\mu}{2}\right)\sin\alpha
    \\
    \cos\alpha
  \end{bmatrix}
\end{align}
Along $C_2$, the azimuthal angle $\varphi+\frac{\mu}{2}$ of $\bvec{n}$ remains constant, while the polar angle $\alpha$ varies according to Eq.~\eqref{eq:alpha_def}.
When $\lambda$ changes by $2\pi$, Appendix~\ref{sec:theta_alpha_vs_lambda} gives
\begin{align}
  \label{eq:alpha_holonomy}
  \alpha(\mu,\beta,\lambda\pm2\pi)
  =\alpha(\mu,\beta,\lambda)\pm\pi
  .
\end{align}
Thus, the adiabatic cycle $C_2$ reverses $\bvec{n}$.

Thus, along $C_2$, the change in the rotation angle $\theta$ given by Eq.~\eqref{eq:theta_holonomy} and the reversal of the rotation axis $\bvec{n}$ occur simultaneously.
Since a rotation matrix satisfies
$\R{\bvec{n}}{\theta}=\R{-\bvec{n}}{2\pi-\theta}$,
the Floquet matrix returns to its initial value, and $C_2$ forms a closed path.

Next, we use the trajectory of the angle vector $\bvec{\theta}$ to show that the trajectory of $\Floquet$ is homotopic to $\gamma$.
Let $\theta_0$ denote the rotation angle at the initial point.
Then, at the final point, $\theta_1=2\pi - \theta_0$.
For the rotation axis, we have $\bvec{n}_1=-\bvec{n}_0$.

For $\theta_0\in (0,\pi)$, we have $\theta_1\in (\pi, 2\pi)$.
Thus, $\theta=\pi$ at some point along $C_2$, and $\bvec{\theta}$ reaches the sphere of radius $\pi$.
Since the initial and final points lie on opposite sides of $\theta=\pi$, the trajectory of $\bvec{\theta}$ along $C_2$ crosses the sphere of radius $\pi$ an odd number of times.
Hence, $\nu(C_2)$ is odd, and the path of $\Floquet$ along $C_2$ is homotopic to $\gamma$.
The case $\theta_0\in (\pi,2\pi)$ is similar.

On the other hand, when $\theta_0=\pi$, $\bvec{\theta}$ starts from the initial point, visits neither the initial point nor its antipodal point along the way, and reaches the antipodal point for the first time at the final point.
Thus, the path traced by $\bvec{\theta}$ is also homotopic to $\gamma$.
Together with the discussion in Sec.~\ref{sec:homotopy}, this result shows that the adiabatic cycle $C_2$ induces an holonomy of the action variable.

We note that the two directions of $C_2$ are not always equally accessible because $\lambda=-\gamma_{S}|\Bp|T$ [Eq.~\eqref{eq:mu_lambda_def}]. In particular, for positive (negative) $\gamma_{S}$, increasing (decreasing) $\lambda$ by $2\pi$ requires decreasing $|\Bp|$ and is possible
only when the initial value of $|\Bp|$ satisfies
$|\Bp|\geq 2\pi/(|\gamma_{S}|T)$.

\section{Discussion}
\label{sec:discussion}

In this paper, we have shown that, for the completely integrable classical map~\eqref{eq:map}, the homotopy class of an adiabatic cycle governs action variable holonomy.
As shown in Sec.~\ref{sec:adevol}, the sign reversal of the rotation axis $\bvec{n}$ of the rotation matrix $\rotmat$ during the adiabatic cycle induces action variable holonomy.

We require the functional form of the action variable to vary continuously during the adiabatic change, which in turn requires $\bvec{n}$ to vary continuously.
Under this requirement, the sign reversal of $\bvec{n}$ can be regarded as a geometric phase factor~\cite{Berry-PRSLA-392-45,Bohm-GPQS-2003,Chruscinski-GPCQM-2004}, because the rotation axis $\bvec{n}$ is a normalized real eigenvector of $\rotmat$ corresponding to the eigenvalue $1$.
Such a sign reversal of a real eigenvector is known as the Longuet-Higgins phase in the study of geometric phases~\cite{LonguetHiggins-PRSL-344-147}.

The rotation axis $\bvec{n}$ of the classical map~\eqref{eq:map} may appear to be a directly controllable parameter. However, the parameter actually controlled is $\rotmat$, while $\bvec{n}$ is determined indirectly as its eigenvector.
Consequently, $\bvec{n}$ can acquire a geometric phase determined by an adiabatic cycle.
This point can be contrasted with the case of an adiabatically varying static magnetic field applied to a classical spin. For a static magnetic field, its direction $\bvec{n}$ itself is an externally controllable parameter. As long as the field magnitude is nonzero, an adiabatic path along which the field direction is reversed between the initial and final points is an open path and does not form an adiabatic cycle.
In this sense, the existence of an adiabatic cycle such as $\gamma$, along which $\bvec{n}$ is reversed, may be regarded as a feature specific to Floquet systems.

Action variable holonomy induces adiabatic torus pumping in phase space.
In this case, the torus corresponding to the initial state of an adiabatic cycle generally does not coincide with that of the final state.
Hence, the condition required to define the Hannay angle is not satisfied. An exception is the torus specified by $I(\bvec{n})=0$.
This torus returns to itself at the end of the cycle regardless of whether the action variable exhibits holonomy.
Therefore, the Hannay angle should be definable in this case.
This situation suggests an analogy with the geometric phase of the $m=0$ state for an open path, as discussed in Ref.~\cite{Robbins-JPA-27-L435} in the study of phase holonomy in quantum systems.

Next, we compare action variable holonomy in the present example with that in the harmonic chain with clamps discussed in Ref.~\cite{Tanaka-PLA-460-128621}. In the latter, action variable holonomy was realized by exploiting crossings of eigenmodes.
Since eigenmode crossings were induced by splitting and reconnecting the system, the adiabatic process there was singular, involving a discontinuous change in the number of degrees of freedom.
In contrast, the present example exploits the Floquet nature of the system and therefore requires no such singular process.

Finally, we mention quantum counterparts of the changes in the action variable and adiabatic torus pumping discussed above, and consider possible directions for theoretical extensions. The adiabatic theorem for bound states in quantum systems is easier to discuss than its classical counterpart because it does not rely on integrability of the dynamics. In quantum systems, while the populations of bound levels are adiabatic invariants, excitation and de-excitation can be induced by adiabatic cycles~\cite{Cheon-PLA-248-285,Tanaka-PRL-98-160407,Cheon-EPL-85-20001,TANAKA-AP-85-1340}.
A topological formulation of these phenomena in terms of fiber bundles has been developed~\cite{Tanaka-PLA-379-1693}. This formulation corresponds to Simon's fiber-bundle formulation of geometric phase factors~\cite{Simon-PRL-51-2167,Bohm-GPQS-2003,Chruscinski-GPCQM-2004}.
In contrast, the homotopy classification of adiabatic cycles discussed in Sec.~\ref{sec:homotopy} is specific to the map~\eqref{eq:map}. Extending this argument to completely integrable systems in general remains a subject for future work.

\section{Summary}
\label{sec:summary}
We investigated action variable holonomy and the associated adiabatic torus pumping in a completely integrable classical Floquet system. The value of the action variable is adiabatically invariant. On the other hand, its functional form on phase space may fail to return to its original form under continuation along an adiabatic cycle and instead change sign. In this case, adiabatic torus pumping in phase space is induced.
We further showed that this phenomenon is characterized by the homotopy class of the adiabatic cycle and explicitly demonstrated it in the classical kicked top.
Thus, not only local adiabatic invariance but also the global topology of parameter space plays an essential role in the adiabatic behavior of the action variable.


\appendix

\section{Derivation of Eq.~\eqref{eq:F3_def}}
\label{sec:eq:F_into_F3}

In rewriting $\Floquet$ defined by Eq.~\eqref{eq:Floquet_def} in the form of Eq.~\eqref{eq:F_into_F3}, we introduced $\Floquet_3(\mu,\beta,\lambda)$. In this section, we derive Eq.~\eqref{eq:F3_def} for $\Floquet_3(\mu,\beta,\lambda)$.

First, Eq.~\eqref{eq:F_into_F3} gives
\begin{align}
  \Floquet_3(\mu,\beta,\lambda)
    =
    {}^{\rm t}R_3\left(\varphi+\frac{\pi}{2}\right)%
    \Floquet%
    R_3\left(\varphi+\frac{\pi}{2}\right)
    .
\end{align}
Next, we rearrange the factors in Eq.~\eqref{eq:Floquet_def} for $\Floquet$.
Since $\bs=\bvec{e}_3$, we have $\R{\bs}{\mu}=R_3(\mu)$.
Also, Eq.~\eqref{eq:bp_parametrized} gives
$\bp=R_{3}({\varphi})R_{2}({\beta})\bvec{e}_3$.
Therefore,
$\R{\bp}{\lambda}%
=R_{3}({\varphi})R_{2}({\beta})R_{3}({\lambda})R_{2}(-{\beta})R_{3}(-{\varphi})%
$.
Thus,
\begin{align}
  &R_3\left(-\varphi-\frac{\pi}{2}\right)\R{\bp}{\lambda} R_3\left(\varphi+\frac{\pi}{2}\right)
    \nonumber\\
  &
    = R_3\left(-\frac{\pi}{2}\right)
    R_{2}({\beta})R_{3}({\lambda})R_{2}(-{\beta})R_{3}\left(\frac{\pi}{2}\right)
    \nonumber\\
  &
  = R_{1}({\beta})R_{3}({\lambda})R_{1}(-{\beta})
    .
\end{align}
Here, we used
\begin{align}
  {}^{\textrm{t}}R_3\left(\frac{\pi}{2}\right)R_{2}({\beta})R_3\left(\frac{\pi}{2}\right)
  = R_1\left(\beta\right)
  .
\end{align}
Eq.~\eqref{eq:F3_def} then follows immediately.

\section{Rotation Axis and Rotation Angle of $\Floquet$}
\label{sec:Floquet_rotation_axis_and_angle}

As explained in the main text, $\Floquet$ (Eq.~\eqref{eq:Floquet_def}) can be regarded as a rotation matrix. In this section, we show that its rotation angle $\theta$ satisfies Eq.~\eqref{eq:theta_eq} and that its rotation axis $\bvec{n}$ is given by Eq.~\eqref{eq:n_def}.

As preparation, we introduce the rotation matrix
\begin{align}
  &
    \Fs(\mu,\beta,\lambda)
    \nonumber\\
  &
    =
    {}^{\rm t}\Rmat{3}{\frac{\mu}{2}}
    \Floquet_3(\mu,\beta,\lambda)
    \Rmat{3}{\frac{\mu}{2}}
    \nonumber\\
  &
  =
  \Rmat{3}{\frac{\mu}{2}}\Rmat{1}{\beta}\Rmat{3}{\lambda}
  \left({}^{\rm t}\Rmat{1}{\beta}\right)
  \Rmat{3}{\frac{\mu}{2}}
\end{align}
obtained by ``symmetrizing'' $\Floquet_3(\mu,\beta,\lambda)$ through a similarity transformation.
Once the rotation axis $\bvec{n}''$ of $\Fs(\mu,\beta,\lambda)$ is found, $\bvec{n}$ is determined by
\begin{align}
  \label{eq:n_by_n_doubleprime}
  \bvec{n}
  =\Rmat{3}{\varphi+\frac{\pi+\mu}{2}}\bvec{n}''
  .
\end{align}

Using the fact that $\bvec{n}''$ is a normalized eigenvector of $\Fs(\mu,\beta,\lambda)$ corresponding to the eigenvalue $1$, we determine $\bvec{n}''$.
First, one eigenvector of $\Fs(\mu,\beta,\lambda)$ corresponding to the eigenvalue $1$ is
\begin{align}
  \tilde{\bvec{n}}''
  =
  \begin{bmatrix}
    0
    \\
    -\sin\beta\sin\frac{\lambda}{2}
    \\
    \cos\frac{\lambda}{2}\sin\frac{\mu}{2}
    +\cos\beta\sin\frac{\lambda}{2}\cos\frac{\mu}{2}
  \end{bmatrix}
\end{align}
This can be verified by direct calculation.

Using $\theta$ satisfying Eq.~\eqref{eq:theta_eq} in the main text, we can normalize $\tilde{\bvec{n}}''$ as
\begin{align}
  \bvec{n}''=\frac{1}{\sin(\theta/2)}\tilde{\bvec{n}}''
  .
\end{align}
However, when $\sin(\theta/2)=0$, we have $\tilde{\bvec{n}}''=\bvec{0}$, which cannot be normalized.
In this case, $\theta=0\mod 2\pi$, and hence
$\Fs(\mu,\beta,\lambda)$, $\Floquet_3(\mu,\beta,\lambda)$, and $\Floquet$
are all equal to
the identity matrix.
The conditions for $\theta=0\mod 2\pi$ are
$\mu=0\mod2\pi$ and $\lambda=0\mod2\pi$,
or
$\beta=0\mod\pi$ and $\mu+\lambda\cos\beta=0\mod2\pi$.

We assume $\theta\ne0\mod 2\pi$ in the following.
Using Eq.~\eqref{eq:alpha_def} in the main text, we obtain
\begin{align}
  \bvec{n}''
  =
  \begin{bmatrix}
    0\\ -\sin\alpha\\ \cos\alpha
  \end{bmatrix}
\end{align}
Together with Eq.~\eqref{eq:n_by_n_doubleprime}, this immediately gives Eq.~\eqref{eq:n_def} in the main text.

\section{$\lambda$ dependence of $\theta$ and $\alpha$}
\label{sec:theta_alpha_vs_lambda}

For given $(\mu,\beta,\lambda)$, we show that $\theta$ and $\alpha$ satisfying Eqs.~\eqref{eq:theta_eq} and \eqref{eq:alpha_def} obey Eqs.~\eqref{eq:theta_holonomy} and~\eqref{eq:alpha_holonomy}.

There are multiple branches of $\theta$ and $\alpha$ satisfying Eqs.~\eqref{eq:theta_eq} and \eqref{eq:alpha_def}.
In particular, in the range $0<\mu<2\pi$, $0<\beta<\pi$, and $0<\lambda<2\pi$, we denote $\theta$ and $\alpha$ on the branch satisfying $0<\theta<2\pi$ and $0<\alpha<\pi$ by $\theta_+(\mu,\beta,\lambda)$ and $\alpha_+(\mu,\beta,\lambda)$, respectively.
We refer to it as the $+$ branch.
Here,
\begin{align}
  \theta_-(\mu,\beta,\lambda)&=4\pi-\theta_+(\mu,\beta,\lambda)
                               ,\\
  \alpha_-(\mu,\beta,\lambda)&=\alpha_+(\mu,\beta,\lambda)+\pi
\end{align}
also satisfy Eqs.~\eqref{eq:theta_eq} and \eqref{eq:alpha_def}, and we refer to them as the $-$ branch.
The rotation axes~\eqref{eq:n_def} corresponding to the $\pm$ branches point in opposite directions.

First, we examine the $+$ branch in the range
$0<\mu<2\pi$, $0<\beta<\pi$, $0 < \lambda<2\pi$.
To use a geometric interpretation based on spherical trigonometry, we rewrite Eqs.~\eqref{eq:theta_eq} and \eqref{eq:alpha_def} as
\begin{align}
  \label{eq:theta_condition_for_sp}
  \cos\frac{\theta}{2}
  &=\cos\frac{\lambda}{2}\cos\frac{\mu}{2}+\cos(\pi-\beta)\sin\frac{\lambda}{2}\sin\frac{\mu}{2}
    ,
  \\
  \label{eq:alpha_condition_cos_for_sp}
  \cos\alpha
  &= \frac{\cos\frac{\lambda}{2}\sin\frac{\mu}{2}%
  -\cos(\pi-\beta)\sin\frac{\lambda}{2}\cos\frac{\mu}{2}}{\sin\frac{\theta}{2}}
    ,
  \\
  \label{eq:alpha_condition_sin_for_sp}
  \frac{\sin(\pi-\beta)}{\sin\frac{\theta}{2}}
  &= \frac{\sin\alpha}{\sin\frac{\lambda}{2}}
    .
\end{align}
\begin{figure}[htb]
  \centering
  \includegraphics[%
        height=5cm%
        ]{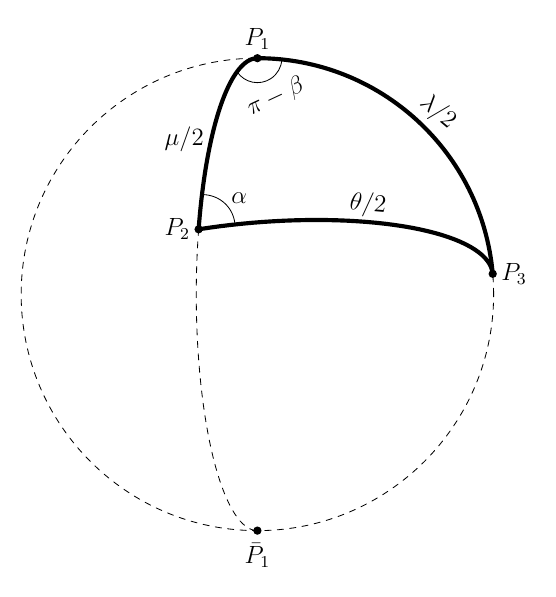} 
  \caption{%
    Spherical triangle $P_1P_2P_3$ (thick lines) on the unit sphere. $\bar{P}_1$ is the antipodal point of $P_1$. Of the two hemispheres bounded by the great circle (dashed line) through $P_1$, $P_3$, and $\bar{P}_1$, the one containing $P_2$ is shown. A part of the great circle through $P_1$, $P_2$, and $\bar{P}_1$ is also shown by a dashed curve.
  }
  \label{fig:spherical_triangle}
\end{figure}
We introduce the spherical triangle $P_1P_2P_3$ on the unit sphere corresponding to these equations (FIG.~\ref{fig:spherical_triangle}):
the arc lengths of the minor arcs $P_1P_2$ and $P_1P_3$ are taken to be $\mu/2$ and $\lambda/2$, respectively.
Furthermore, the angle $\angle P_1$, i.e., the angle between the tangents at $P_1$ to the minor arcs $P_1P_2$ and $P_1P_3$, is taken to be $\pi-\beta$.
Then, Eq.~\eqref{eq:theta_condition_for_sp} is the spherical law of cosines for this triangle, and Eq.~\eqref{eq:alpha_condition_cos_for_sp} can be regarded as a consequence of the spherical law of cosines.
We take the arc length of the minor arc $P_2P_3$ to be $\theta/2$ and the angle $\angle P_2$ to be $\alpha$.
Then, Eq.~\eqref{eq:alpha_condition_sin_for_sp} is the spherical law of sines for this triangle.

We now describe $\theta$ and $\alpha$ shown in Fig.~\ref{fig:spherical_triangle}.
For given $(\mu,\beta,\lambda)\in (0,2\pi)\times(0,\pi)\times(0,2\pi)$, this spherical triangle is uniquely determined.
Since this spherical triangle lies within a hemisphere, there exists a unique pair $(\theta,\alpha)$ satisfying $0 <\theta<2\pi$, $0 <\alpha<\pi$.
This is the $+$ branch.

Next, we examine the $\lambda$ dependence of the $+$ branch using FIG.~\ref{fig:spherical_triangle}.
For $0 < \lambda < 2\pi$, $\alpha$ increases monotonically with $\lambda$.
Using the one-sided limit $\lambda\to0+$, we set $\theta_+(\mu,\beta,0)=\mu$ and $\alpha_+(\mu,\beta,0)=0$.
Using the other one-sided limit $\lambda\to2\pi-$, we set $\theta_+(\mu,\beta,2\pi)=2\pi-\mu$ and $\alpha_+(\mu,\beta,2\pi)=\pi$.

From the above results, the properties of the $-$ branch for
$\mu\in(0,2\pi)$, $\beta\in(0,\pi)$, $\lambda\in(0,2\pi)$ also follow.
Its ranges are
$2\pi <\theta_-(\mu,\beta,\lambda)<4\pi$,
$\pi <\alpha_-(\mu,\beta,\lambda)<2\pi$,
and $\alpha_-$ increases monotonically with $\lambda$.
Furthermore, the values at $\lambda=0, 2\pi$ are
$\theta_-(\mu,\beta,0)=4\pi-\mu$,
$\alpha_-(\mu,\beta,0)=\pi$,
$\theta_-(\mu,\beta,2\pi)=\mu + 2\pi$,
$\alpha_-(\mu,\beta,2\pi)=2\pi$.

Using the above results, we can extend the range of $\lambda$ while keeping
$\mu\in(0,2\pi)$ and $\beta\in(0,\pi)$.
By comparing the $+$ branch at $\lambda=2\pi$ with the $-$ branch at $\lambda=0$, we find that the $+$ branch for $2\pi \le \lambda \le 4\pi$ is given by
\begin{align}
  \theta_+(\mu,\beta,\lambda)
  &= \theta_-(\mu,\beta,\lambda-2\pi)-2\pi
  ,\\
  \alpha_+(\mu,\beta,\lambda)
  &=
  \alpha_-(\mu,\beta,\lambda-2\pi)
  .
\end{align}
Therefore,
\begin{align}
  \label{eq:plus_shift_lambda_by_2pi}
  \theta_+(\mu,\beta,\lambda)
  &
    =2\pi - \theta_+(\mu,\beta,\lambda-2\pi)
  \\
  \alpha_+(\mu,\beta,\lambda)
  &
    =\alpha_+(\mu,\beta,\lambda-2\pi)+\pi
    .
\end{align}
Extending the $+$ branch in the same way and applying the result to the $-$ branch, we obtain Eqs.~\eqref{eq:theta_holonomy} and~\eqref{eq:alpha_holonomy}.

%

\end{document}